\documentstyle[aps,epsfig,multicol,fancyheadings]{revtex}
\begin{document}

\title{A Cellular Automata Model for Citrus Variagated Chlorosis}

\author{M. L. Martins, G. Ceotto, S. G. Alves, C. C. B. Bufon, J. M. Silva, and F. F. Laranjeira$^{*}$ }
\address{Departamento de F\'{\i}sica, Universidade Federal de Vi\c cosa, 36571-000, 
Vi\c cosa, Brazil;}
\address{$^{*}$ Instituto Agron\^omico de Campinas,13001-970, Campinas, Brazil;}

\maketitle

\begin{abstract}
A cellular automata model is proposed to analyze the progress of Citrus Variegated 
Chlorosis epidemics in S\~ao Paulo oranges plantation. In this model epidemiological 
and environmental features, such as motility of sharpshooter vectors which perform L\'evy 
flights, hydric and nutritional level of plant stress and seasonal climatic effects, are included. The observed epidemics data were 
quantitatively reproduced by the proposed model varying the parameters controlling vectors motility, plant stress and initial population 
of diseased plants.
\end{abstract}

\vspace{1cm}
\vspace{1cm}

\begin{multicols}{2}[]

\narrowtext

\section{Introduction}

Citrus Variegated Chlorosis (CVC) is an economically relevant disease affecting citrus ~\cite{Chico}. In the S\~ao Paulo region (Brazil),
 one of the important citrus growing areas of the world, responsible for about $30\%$ of the world production, CVC reduces the size and 
number of fruits by more than $35\%$~\cite{Palazzo}. CVC is considered to be potentially the most devastating citrus disease and 
represents the main threat to the Brazilian citrus industry, with annual revenues of the order of $1.2$ to $1.4$ billions of dollars. 
The losses associated with the disease are estimated in about $100$ millions dollar yearly~\cite{Chico}.

CVC is caused by a xylem-limited bacterium, {\it Xyllela fastidiosa}~\cite{Beretta}, transmitted  by xylem feeding, suctorial  
sharpshooter leafhoppers (Hemiptera: Cicadellidae)~\cite{Lopes,Roberto}. In S\~ao Paulo, the species {\it Dilobopterus costalimai} 
appears to be the most efficient vector for CVC transmission~\cite{Roberto}. Nowadays, a sweet orange cultivar resistant to {\it X. 
fastidiosa} is unknown, and control practices for CVC (bactericidal agents, systematic pruning of infected branches, chemical control 
of vectors, and/or rouging of severe affected plants) are expensive, ineffective, or environmentally damaging.

Recent studies on various aspects of the epidemiology of CVC (\cite{Chico}, and references 
therein) have provided fundamental information which can be used to develop a cellular 
automata (CA) model of the pathosystem. CA or other epidemic models could become relevant tools to address numerous practical and 
experimental questions: forecasting the progress and final intensity of CVC, planning and evaluation of strategies for disease
 control and determination 
of the relevant mechanisms involved in the disease spreading.

In this paper, we propose a simple CA model to simulate the CVC progress in which some epidemiological and environmental features, 
such as vectors motility, plant stress and 
seasonal modulations are included. The simulational results are compared with the CVC 
progress curves in time and spatial infection patterns observed in the S\~ao Paulo region.

\section{Experimental data on CVC epidemics}

\subsection{CVC progress in time}

The CVC epidemic was observed by visual assessments of typical symptoms occurring on leaves or fruits, in eleven groves of P\^era, 
Hamlin and Natal sweet oranges cultivated in two farms of the northern areas (Bebedouro and Colina counties) of S\~ao Paulo state, 
Brazil. In such areas, severely attacked by CVC, are planted the more susceptible cultivars having the supposedly more propitious 
age for disease development. The field data were collected over a 20-month period, from September, 1994 through March, 1996. The 
CVC incidence were bimonthly mapped and the data for each area and each evaluation were transformed to proportion of symptomatic 
plants for temporal characterization of the disease spreading.

\begin{figure}[f]
\centerline{\epsfig{file=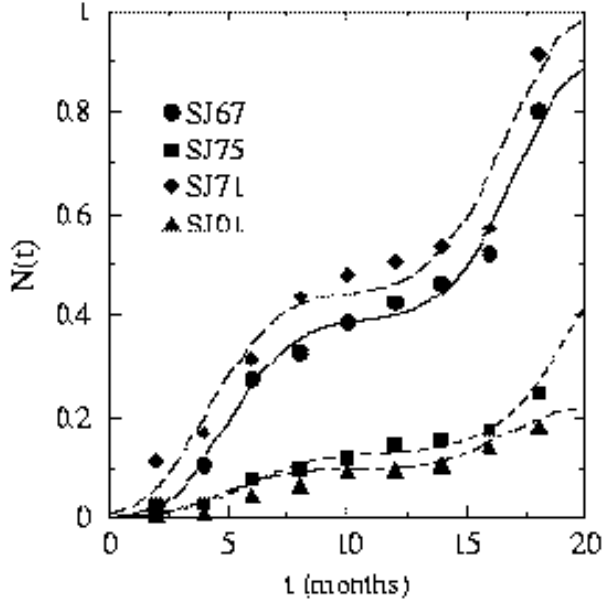,width=9cm,angle=0}}
\caption{Observed CVC progress curves in four groves SJ01, SJ67, SJ71 and SJ75. The curves represent 
generalized five parameter logistic fittings to the observed data.}
\label{fig1}
\end{figure}

The CVC progress curves are shown, for four different groves, in figure \ref{fig1}. All of them are 
double sigmoid, which is a clear indication that CVC is a polycyclic disease characterized 
by the existence of two phases: one in which the disease spreading is fast, contrasted by 
another one in which the epidemic development is almost stopped. For each grove the observed 
data sets are fitted by five parameter logistic curves~\cite{Hau} of the form

\begin{equation}
N(t)=\frac{p_1}{1+exp[-(p_2+p_3t+p_4t^2+p_5t^3)]}.
\end{equation}

Table 1 gives the corresponding parameters and the coefficients of determination ($R^2$) 
have been listed. In addition to the $R^2$ coefficient, the residual sums of squares for 
error and the consistency of the predict values for the upper asymptotic fraction of diseased plants ($p_1\leq 1$) were 
take into account to select the logistic among the Gompertz and 
monomolecular generalized models with four or five parameters. It is important to notice that sigmoid curves can be 
generated by distinct models (fitting equations). Indeed, the temporal increase of citrus tristeza virus, whose 
vectors are aphid species, follows a nonlinear Gompertz model~\cite{Gottwald}.

\subsection{Infected cluster size distribution functions.} 

On any one of the orangeries containing about $10^3$ trees there is no unique inoculum source. Each single infected 
plant or small initial group of infected plants grows by inoculating its adjacent neighbors, and aggregates with 
other affected trees, forming large clusters. As a result, the mean cluster size of infected trees increases in 
time and, in order to describe the disease spreading, it is necessary to investigate the dynamic aspects of the 
distribution of infected plant aggregates generated by CVC progress.

\begin{figure}[f]
\centerline{\epsfig{file=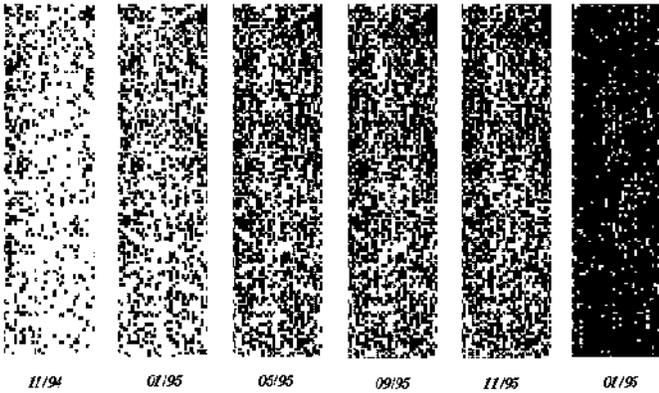,width=9cm,angle=0}}
\caption{Temporal sequence of CVC incidence maps in the grove SJ 71 with 3960 P\^era orange trees. 
Six representative evaluations (11/94, 01/95, 05/95, 09/95, 11/95 and 01/96) 
are shown. Each black square corresponds to one symptomatic plant.}
\label{fig2}
\end{figure}

A cluster of diseased plants have been defined as any set of interconnected infected trees which are spatially 
isolated from any other group of diseased plants in the orangery. Then, the cluster size distribution function 
$n_s(t)$ is the fraction of these clusters consisting of $s$ infected plants at time $t$, and were directly 
obtained by counting the number and size of diseased clusters present in the spatial patterns of CVC epidemic, 
like those shown in figure \ref{fig2}, at each observation time.

\begin{figure}[f]
\centerline{\epsfig{file=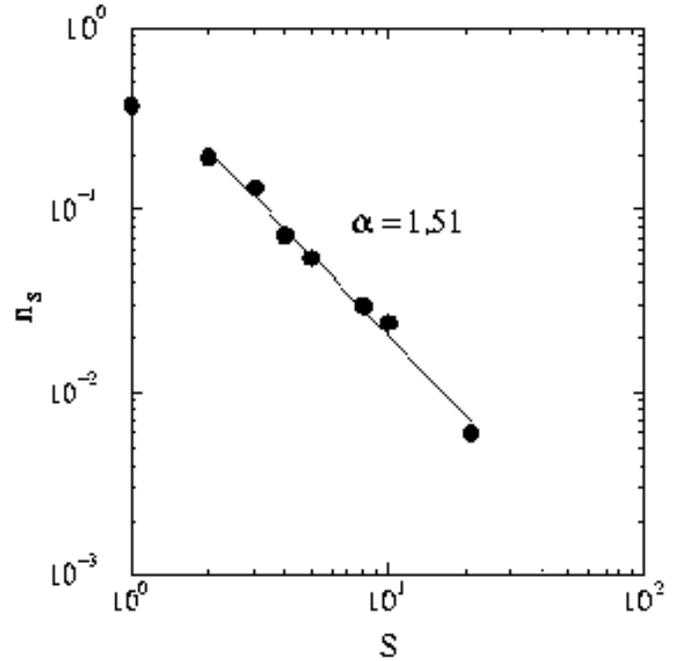,width=9cm,angle=0}}
\caption{Observed cluster size distribution function $n_s(t)$ of clusters containing $s$ diseased plants 
for the grove SJ71 at the observation time $t=6$ months. The straight line represent the best fit to the data; 
its slope gives the exponent $\alpha$ describing the power-law decay of $n_s(t)$.}
\label{fig3}
\end{figure}

Figure \ref{fig3} shows the distribution $n_s(t)$ for one of the observed orangeries. It suggests that 
$n_s(t)$ follows a power-law distribution, that is, $n_s(t)\sim s^{-\alpha }$, over at least 
one decade of their argument $s$ at any given time of CVC progress. A power-law distribution indicates the 
absence of a characteristic scale for the size of diseased plant aggregates. The exponents describing the 
power-law decay of $n_s(t)$ have values between $1.4$ and $1.8$ for all the studied orangeries. These 
values are characteristic of an $1/f$ noise~\cite{Bak}, as are commonly referred the $1/f^{\gamma}$ 
spectra with $\gamma$ in the interval $[0,2]$. Therefore, the CVC infection dynamics has an $1/f$-like 
signal, which, from the physical point of view, is a sign of a cooperative phenomenon occurring in a 
spatially extended nonequilibrium system.

\subsection{Self-affine profiles in CVC evolution patterns.}

Self-affine profiles~\cite{Vicsek,Barabasi} can be generated from the spatial patterns of diseased 
plants using various methods. The simplest of them is the 1:1 mapping between a given spatial 
configuration at time $t$, such as those shown in figure \ref{fig2}, and a ``walk process''~\cite{Peng,Stanley}. 
In this method each binary symbol $\sigma_i(t)$, describing the plant state ($\sigma_i(t)=0$: normal; 
$\sigma_i(t)=1$: diseased) is identificated with a step (to the right or to the left) of a one-dimensional walk.

\begin{figure}[f]
\centerline{\epsfig{file=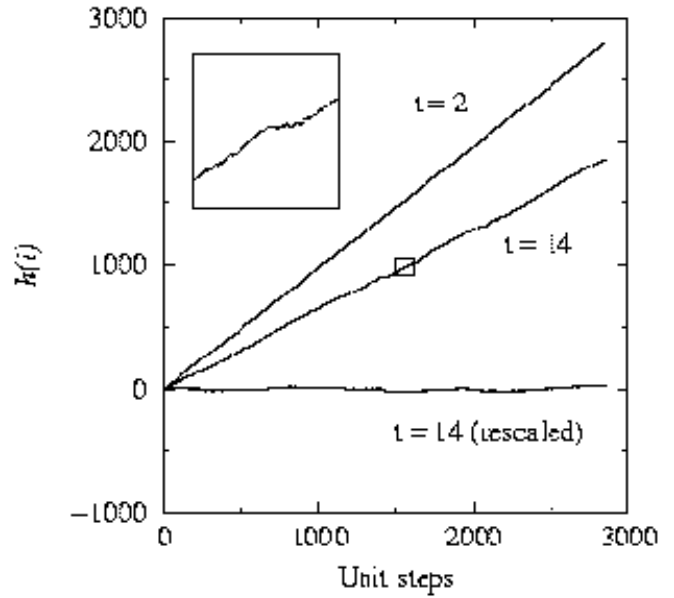,width=9cm,angle=0}}
\caption{Self-affine profiles generated by the walk process method at two distinct 
observation times in the grove SJ75, containing 2,880 P\^era orange plants. The rescaled
profile at $t=14$ months was obtained by subtracting the corresponding linear fitting of 
the data. The inset is an enlargement of the central region of the profile at $t=14$.}
\label{fig4}
\end{figure}

Specifically, to  an unique spatial pattern $\{ \sigma_1(t),\sigma_2(t),\ldots,\sigma_N(t) \}$ of $N$ plants at 
fixed time $t$, corresponds a profile given by the sequence of the walker displacements $h_i$ after $i$ unit 
steps, $\{ h_1(t),h_2(t),\ldots,h_N(t) \}$, defined as

\begin{equation}
h_i(t)=\sum_{j=1}^i\rho _j(t)
\end{equation}
where $\rho _j=1$ if $\sigma _j=1$ (step to the right), or $\rho _j=-1$ if
$\sigma _j=0$ (step to the left). Profiles generated at two distinct observation times in a
given orangery using this walk process are shown in figure \ref{fig4}.

Obtained the profiles by the walk process, we can investigate the nature of their correlations 
through the analysis of the profile roughness~\cite{Barabasi}. The statistical measure $W$, 
which characterizes the roughness of the walk profile, is defined by the rms fluctuation in the displacement

\begin{equation}
W(N,t)= \sqrt{\frac{1}{N} \sum_{i=1}^{N} [h_i(t)-\overline{h}(t)]^2}
\end{equation}
where

\begin{equation}
\overline{h}(t)= \frac{1}{N} \sum_{i=1}^{N} h_i(t)
\end{equation}
is the mean displacement of the walk.

For self-affine profiles the roughness $W(N)$ will be described by a power law scaling

\begin{equation}
W(N) \sim N^H
\end{equation}
with the exponent $H$ restricted to the interval $[0,1]$ and related to the fractal dimension 
of the profile~\cite{Vicsek}. $H=1/2$ corresponds to a random walk; $H>1/2$ implies that the 
profile presents persistent correlations, and profiles with $H<1/2$ are anticorrelated. 

As one can see in figure \ref{fig4}, the profiles generated from the CVC epidemic patterns usually 
have drifts. This is the reason why we use the method of roughness around the rms straight 
line~\cite{Zeh} to evaluate the Hurst exponent $H$. In this method the roughness 
$W(N,\epsilon )$ in the scale $\epsilon $ is given by

\begin{equation}
W(N,\epsilon)=\frac{1}{N} \sum_{i=1}^{N} w_i(\epsilon)
\end{equation}
and the local roughness $w_i(\epsilon)$ is defined as

\begin{equation}
w_i(\epsilon )=\sqrt{\frac{1}{2\epsilon +1}\sum_{j=i-\epsilon }^{i+\epsilon
}\{h_j-[a_i(\epsilon )x_j+b_i(\epsilon )]\}^2}.
\end{equation}
$a_i(\epsilon )$ and $b_i(\epsilon )$ are the linear fitting coefficients to the 
displacement data on the interval $[i-\epsilon ,i+\epsilon ]$ centered on the site 
$i$. Again, self-affine profiles satisfy the scaling law

\begin{equation}
W(\epsilon) \sim \epsilon^H
\end{equation}

The method described here was used to characterize the spatio-temporal patterns generated 
by elementary one-dimensional deterministic cellular automata~\cite{Sales}.

\begin{figure}[f]
\centerline{\epsfig{file=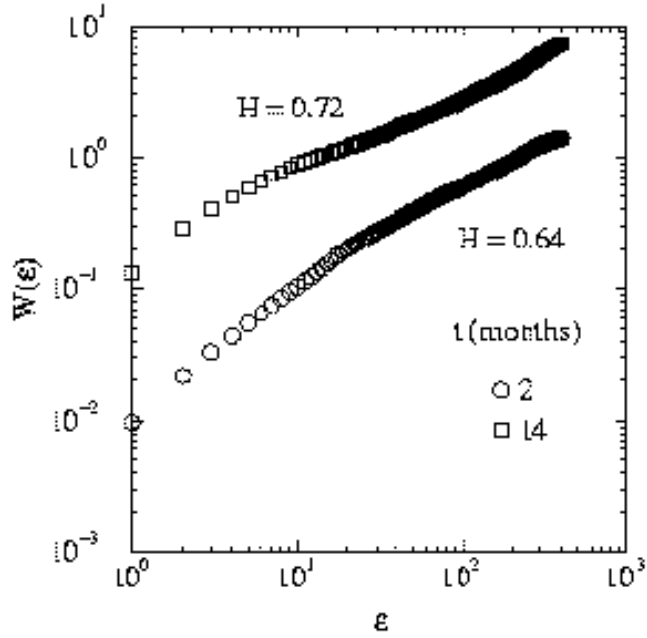,width=9cm,angle=0}}
\caption{Typical log-log plot of $W(\epsilon) \times \epsilon$ used to calculate the 
roughness or Hurst exponent characterizing a rough profile, in this case the CVC spatial 
patterns in grove SJ75 at two distinct observation times. The Hurst exponent corresponds 
to the slope of the straight-line fitting the linear part of the curve.}
\label{fig5}
\end{figure}

A Typical log-log plot of $W(\epsilon )\times \epsilon $ used to calculate the Hurst exponent 
is shown in figure \ref{fig5}. This exponent characterizes the spatial patterns generated by CVC epidemic. 
For all the analyzed orangeries the profiles are self-affine with Hurst exponents markedly distinct 
from $1/2$, which means long-range correlations present in the spatial disease patterns. Also, the 
roughness exponents increases from a initial value around $1/2$, indicating a random infection pattern 
at the beginning, towards its maximum value $1$ which corresponds to a totally infected orangery. Thus 
our results show that CVC epidemic gives rise to aggregated patterns in which the inoculum level tends 
to be high in scattered sequences of neighboring plants, i. e., infected plants tends to be close to another 
diseased trees and the same holds for normal plants. Therefore, it appears that the pathogen occurs in small 
clusters which progressively expands by a contagion process mediated by vectors which predominantly spreads 
from plant to plant.

\section{A Cellular Automata model for CVC}

Stochastic CA models were used before in plant pathology to simulate diseases spreading through spore dispersal~\cite{Minogue},
infection dynamics of {\em R. Solani}~\cite{Kleczkowski} and infection of cereal roots by the take-all fungus~\cite{Gilligan1,Gilligan2},
but traditionally the mathematical modeling of plant disease is based on systems of differential equations~\cite{Gilligan2}.

CAs are totally discrete dynamical systems (discrete space, discrete time and discrete number 
of states) which provide simple models for a great number of problems in science~\cite{Wolfram,Ermentrout}. 
With each site (noted $i$), is associated a variable $\sigma_i$, which can be in $K$ different states 
$\sigma_i=0,1,\ldots,K-1$. The dynamics is defined, at each time step, by rules depending on the values 
at previous time of \{$\sigma_i$\} associated with a given number of $q$ arbitrary sites (called inputs). 
Usually one considers regular lattices and the inputs refer to the sites on the local neighborhood only. 
The local rules of a CA may be probabilistic or deterministic and the sites are simultaneously updated.

In order to design the CA model for CVC spreading, we take into account the following basic 
features of CVC pathosystem characterized in the previous section. The bacterium {\it X. fastidiosa} 
is transmitted by sharpshooter vectors of rather limited motility in the groves. This hypothesis is 
coherent with the results obtained for the roughness or Hurst exponents describing the CVC infection 
patterns. In diseased plants, the bacterium is systemic, nutritional unbalance and general weakness are 
commonly observed. Thus, the infected sites continuously act as inoculum sources to other healthy plants. 
Since there was no measured effect of wind direction or machine based cultural practices on CVC 
spreading~\cite{Chico}, the vectors flies were assumed to be completely random in our model. Also, 
in healthy stressed trees the observed sharpshooter population is small, since these insects are preferable 
attracted by plants with new vegetative growth. Finally, as shown by the CVC progress curves, seasonal effects 
play a central role in disease spreading. In fact, the fastest CVC spreading progress is observed from 
September/October (flowerage) through March (end of summer), a period associated to high temperatures, 
regular rains and vegetative growth. The seasonal modulations are included in the CA model through the 
variation in the motility of sharpshooter vectors as well as in the fraction of normal plants under 
hydric and nutritional stress. The functional form assumed to model such seasonal variations in our 
model is a sine wave function.

In our CA model the orangery is represented by square lattices of linear size $L$ with null 
fixed boundary conditions (isolated grove, i. e., all the state variables are zero outside 
the lattice). The state of each plant is described by two binary variables: $\sigma _{i,j}^{(1)}$ 
and $\sigma _{i,j}^{(2)}$, where $i,j=1,2,...,L$. $\sigma_{i,j}^{(1)}=0$ represents a 
healthy (normal) plant and $\sigma _{i,j}^{(1)}=1$ a diseased plant. On the other hand, 
$\sigma_{i,j}^{(2)}=1$ represents a plant under hydric or nutritional stress and 
$\sigma _{i,j}^{(2)}=0$ a non-stressed plant. Finally, the fraction of inoculative vectors in each 
site is also represented by a binary variable $\tau _{i,j}$. $\tau_{i,j}=0$ means a low 
fraction of inoculative sharpshooter leafhoppers in the insect population  and $\tau_{i,j}=1$ 
a high inoculative fraction.

In all simulations, random initial conditions in which any plant of the lattice is diseased, 
with probability $p_{inf}$, and stressed, with probability $p_{st}(0)$, have been used. Since 
CVC causes severe stress in a diseased plant~\cite{Chico}, a infected tree 
($\sigma _{i,j}^{(1)}=1$) becomes immediately stressed ($\sigma _{i,j}^{(2)}=1$) in our model. 

\begin{figure}[f]
\centerline{\epsfig{file=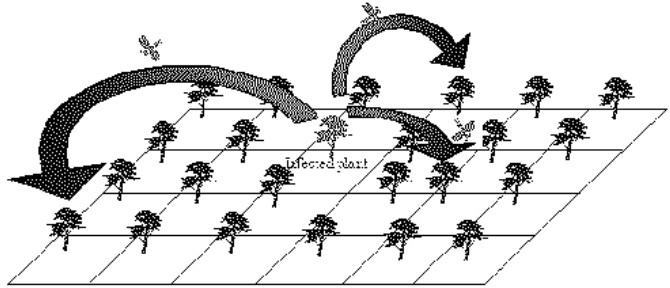,width=9cm,angle=0}}
\caption{Schematic representation of CVC spreading from a diseased plant (central gray square) used in our CA model. 
The central infected site act as an inoculum source for 
$n_v(t)=4$ other distinct sites whose distances $r$ are chosen from a symmetric L\'evy distribution. These ``target'' 
sites are reached through the vectors flies represented by the arrows. $n_v(t)$ change seasonally with time.}
\label{fig6}
\end{figure}

All the sites are simultaneously updated using the following local rules:

i) For an infected site the correspondent state at the next time step is 
$\sigma _{i,j}^{(1)}=1$, $\sigma _{i,j}^{(2)}=1$ (diseased and stressed tree) and 
$\tau_{i,j}=1$ (high fraction of inoculative vectors). In addition, as shown in figure \ref{fig6},
 each infected site acts as an inoculum source for $n_v(t)$ distinct plants at distance 
$r_k$, $k=1,2,\ldots,n_v$, chosen at random accordingly a symmetric L\'evy distribution

\begin{equation}
p(r)=\frac{1}{2\pi}\int_{-\infty}^{+\infty}dt\,\,exp(it(\mu-r)-|t|^a).
\end{equation}
Thus, the lengths of each one of the $n_v(t)$ vector flights are not constant but rather are 
chosen from a probability distribution with a power-law tail. Each one of these selected 
neighbors will assume, at the next time step, the state $\sigma _{i,j}^{(1)}=1$, $\sigma _{i,j}^{(2)}=1$ 
and $\tau_{i,j}=1$, if it is a normal and non-stressed plant. Otherwise, the selected neighbor will stay 
in the same previous state. Thus, in our CA model a healthy stressed tree is not infected by CVC, since 
their main vectors ({\it Dilobopterus Costalimai} and {\it Acrogonia sp.}) are preferentially observed 
in plants exhibiting young buds and leaves. In contrast, a healthy and non-stressed plant becomes 
diseased if it is reached by inoculative vectors coming from at least one infected site.

ii) For a normal (stressed or non-stressed) plant not target by a given diseased plant, the 
correspondent state at the next time step is $\sigma _{i,j}^{(1)}=0$ (normal) and 
$\tau_{i,j}=0$ (low fraction of inoculative vectors). Yet, the value of $\sigma _{i,j}^{(2)}=0$ 
or $1$ is chosen at random with probability $p_{st}\equiv 1-p_{nst}(t)$.

iii) In order to simulate the seasonal effects on both, plant stress and vectors motility,  
the number of inoculative vector flights $n_v(t)$, and the probability associated to a 
non-stressed plant $p_{nst}(t)$ are periodic functions of time given by: 
\begin{equation}
n_v(t)=n_0 +INT[n_0 \,sin(\frac{2\pi t}{T}+\phi)];
\end{equation}
\begin{equation}
p_{nst}(t)=p_{min}+(1-2p_{min})\times \frac{1 +sin(\frac{2\pi t}{T}+\phi)}{2},
\end{equation}
where INT means the integer part, $n_0$ is the time average number of inoculative vector 
flights, $p_{min}$ is the minimum probability to find a non-stressed plant, and $T$ is 
the period of one year. $\phi$ is a common phase angle to describe possible time shifts 
between the simulated and field data. 

At last, we shall discuss some simplifications of our CA model for CVC progress. The use 
of a symmetric periodic function to model the seasonal effects should be thought as a rough 
approximation to the much more complex climatic variations observed in nature. Another simplification 
is that once infected a given plant immediately acts as an inoculum source, 
in contrast to the classical notion of a discontinuous infectious period. It is known that 
the spread of disease involves the interplay of two dynamical processes: the mechanisms of 
transmission and the evolution of the pathogen within hosts. The basic questions on how the 
bacterium spread within the xylem system and what is the mechanism of pathogenesis in CVC are 
unanswered. Particularly, it seems that the CVC symptoms in plants depend on the rate and 
extent of colonization by the bacteria. Also, a recent study~\cite{Hill} shows that 
an efficient transmission of {\it Xylella fastidiosa} by vectors occurs only after its 
population overcome a threshold in plant hosts. In addition, the transmission rate increases 
as the bacterial population in a plant increases. These features are not included in our CA model 
due to the lack of information about the system {\it Xylella fastidiosa} in citrus.

\section{Results}

\begin{figure}[f]
\begin{multicols}{2}[]
\narrowtext
\centerline{\epsfig{file=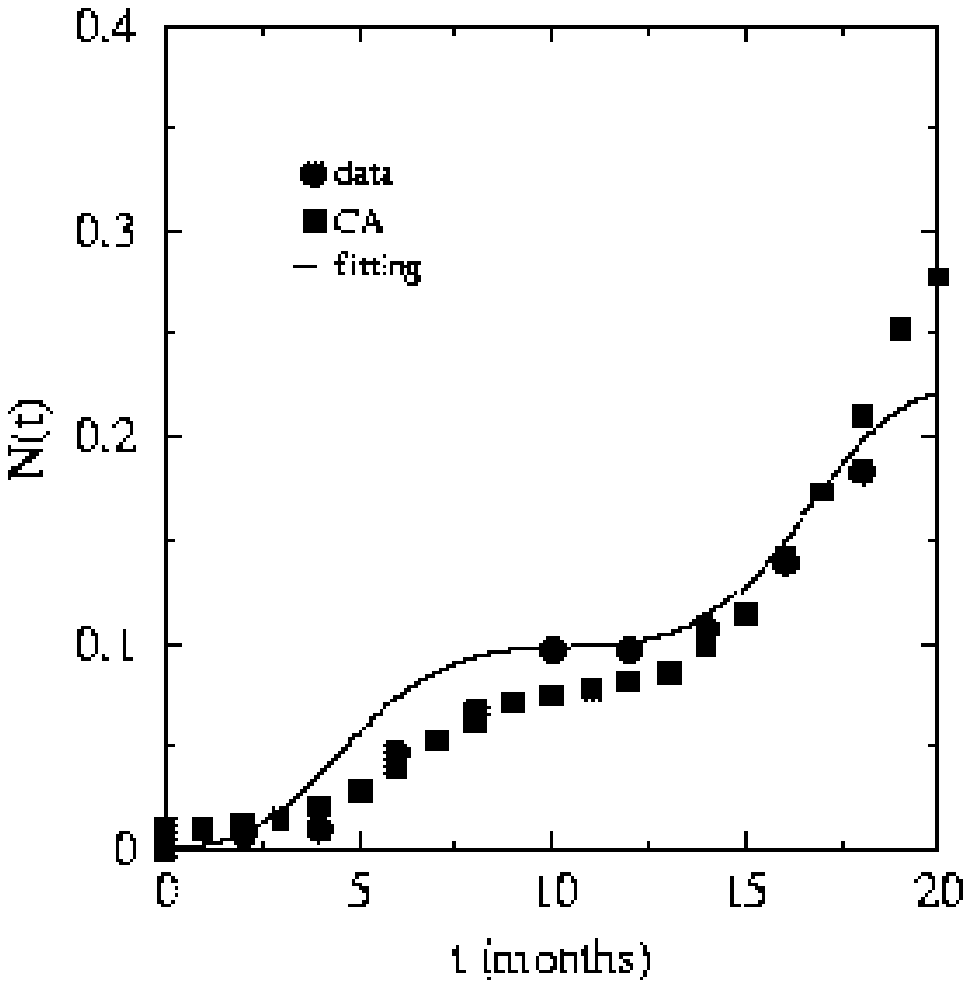,width=7.5cm,angle=0}}
\centerline{\epsfig{file=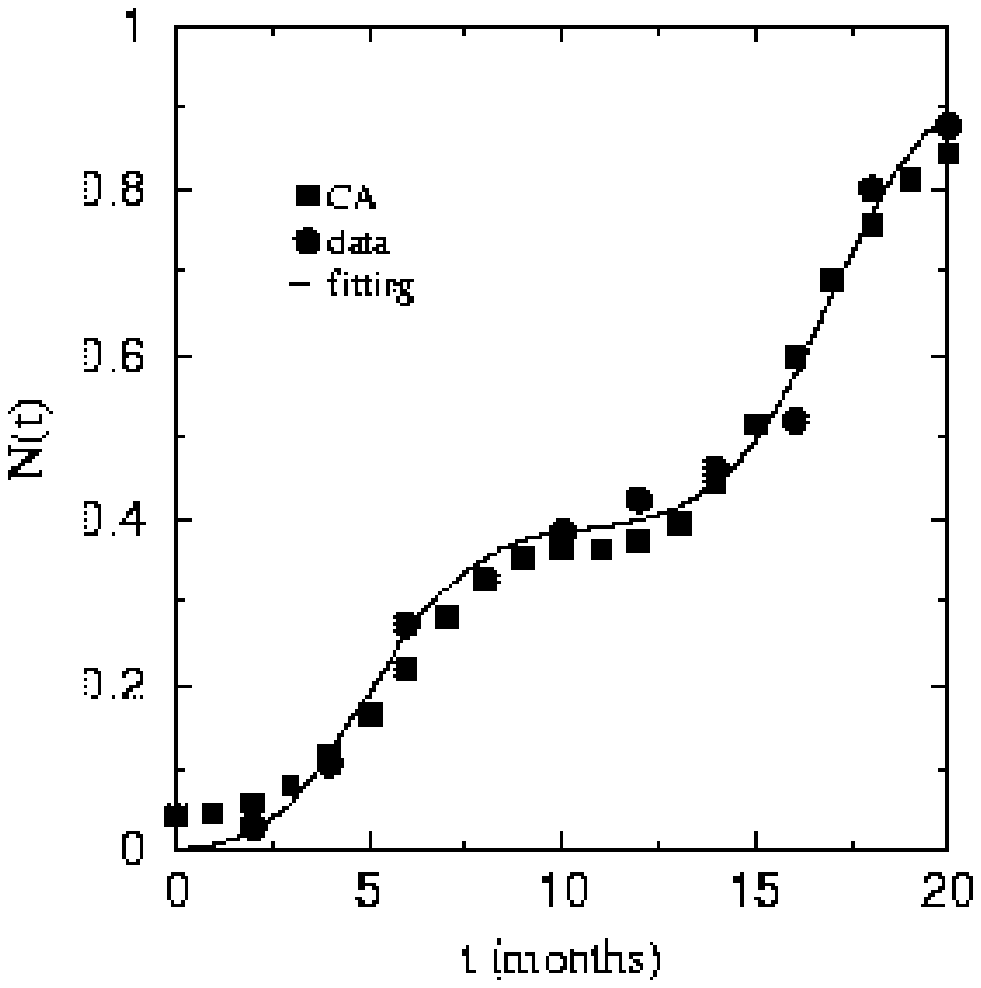,width=7.5cm,angle=0}}
\centerline{\epsfig{file=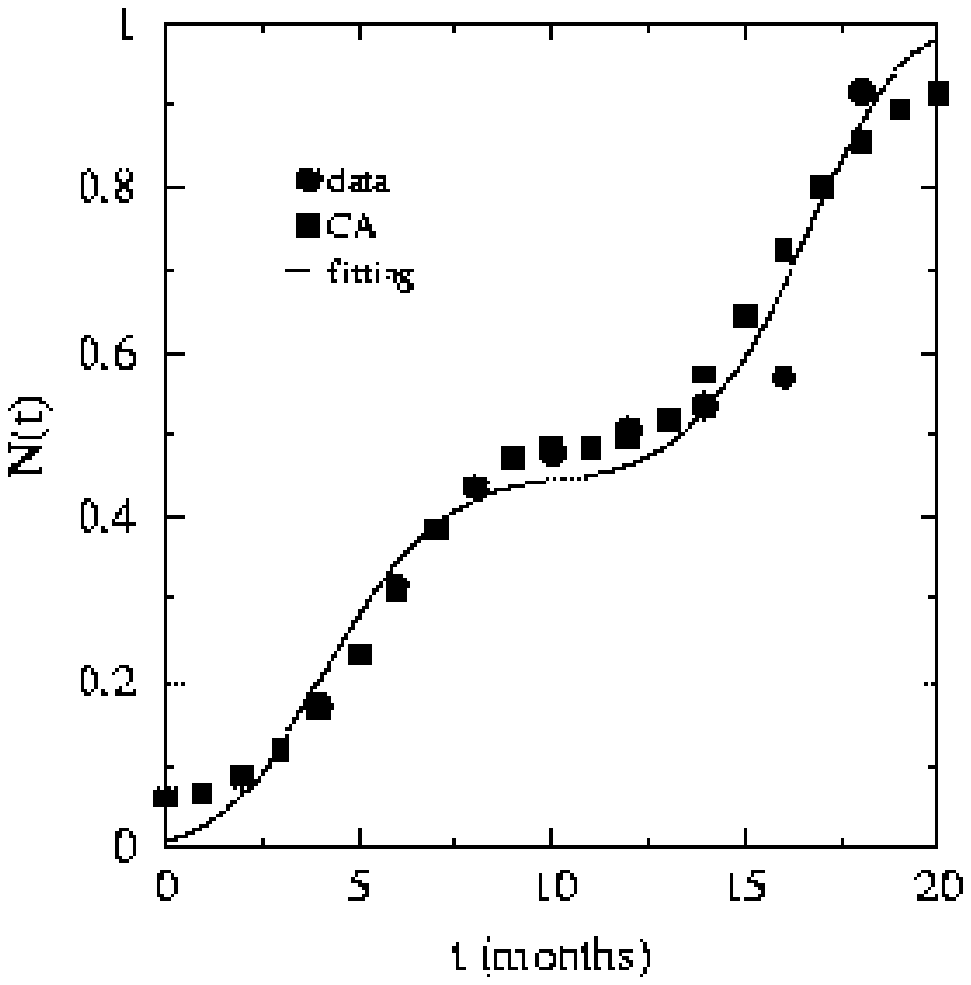,width=7.5cm,angle=0}}
\centerline{\epsfig{file=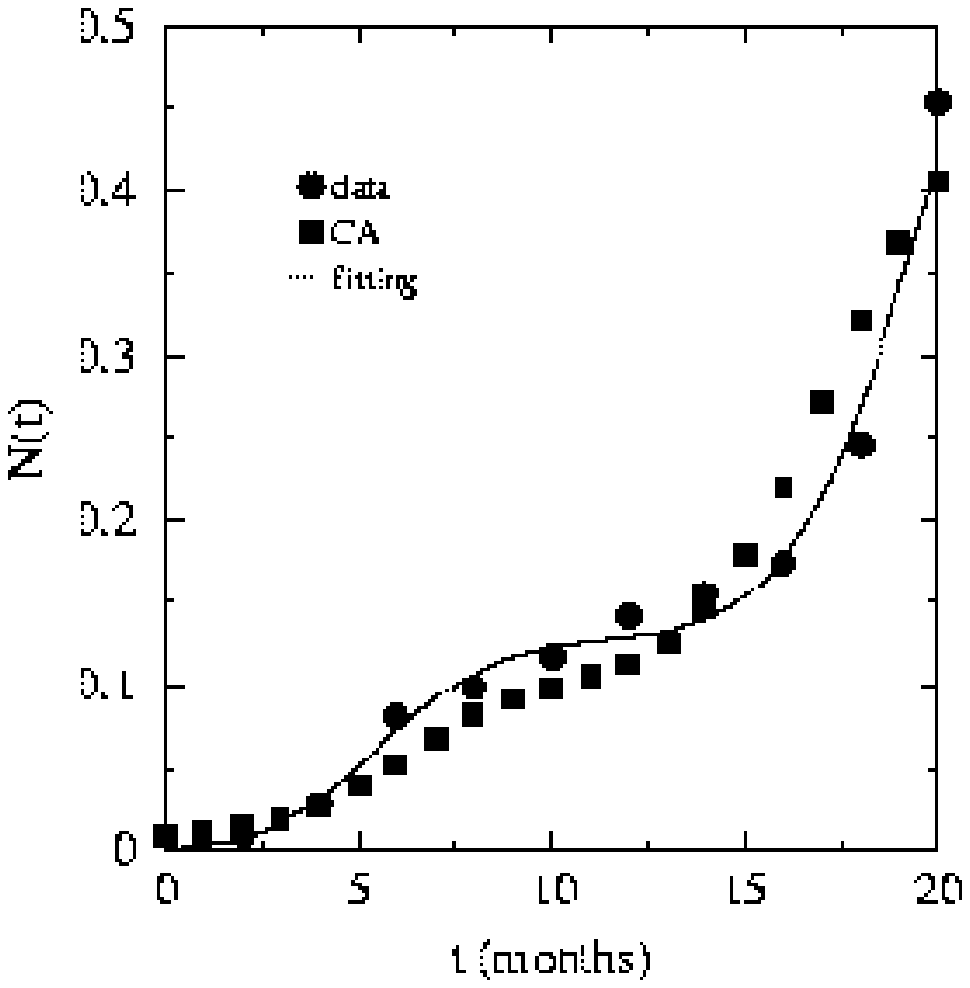,width=7.5cm,angle=0}}
\end{multicols}
\widetext
\caption{Comparison between the CA generated and fitted CVC progress curves for groves
(a) SJ01, (b) SJ67, (c) SJ71 and (d) SJ75. The observed data are also shown, and the used CA parameters are those 
listed in Table 2. The data correspond to an average over $100$ different realizations of CA evolution. The error 
bars are equal or smaller than the symbol syze.}
\label{fig7}
\end{figure}

Now we shall report on the simulational results for our CA model. In all simulations a linear size 
$L=200$, a seasonal period $T=12$ months (one year) and a phase angle $\phi =-60^o$ (a 
shift of two months between the observed and the simulation initial times) were used. The remaining 
five CA parameters, namely, $n_0$, $p_{inf}$, $p_{min}$, $\mu$, and $a$ were varied in 
order to compare the simulated and observed CVC progress curves. The results for four groves 
are shown in figure \ref{fig7}. As one can see, the qualitative behavior of the CA progress curves show 
the same functional behavior of the measured curves. Even a surprisingly quantitative agreement between 
the CA and the observed CVC progress curves was obtained. The CA parameters values are listed in 
Table 2. Therefore our simulation results suggest that the L\'evy distribution of vectors flights 
is universal, with $\mu=4$ and $a=0.68$. The same holds true for $n_0$, the time average number of 
sharpshooter vectors flies starting from each infected plant, which assumed a value $5$ for all four simulated groves.

At this point, it is interesting to note that our CA model predicts even triple, quadruple 
or superior sigmoid progress curves depending on the time elapsed up to a total infection of 
the grove. This simulational prediction can be easily tested simply by observing the CVC progress 
in field for a period greater than $20$ months as done in the present work. Moreover, such triple 
or quadruple sigmoid curves could be mathematically modeled by using generalized logistic, 
monomolecular or Gompertz functions only if the disease progress curves were subdivided into 
three or four parts that are analyzed separately. However, this approach is inadequate to 
describe the entire disease dynamics and to determine several parameters of epidemiological importance~\cite{Amorim1,Amorim2}.

\begin{figure}[f]
\centerline{\epsfig{file=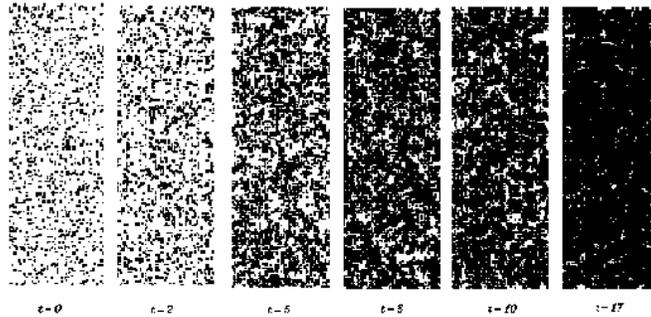,width=9cm,angle=0}}
\caption{A simulated sequence of CVC incidence maps in the grove SJ71. Each black square corresponds to one symptomatic plant.}
\label{fig8}
\end{figure}

\begin{figure}[f]
\centerline{\epsfig{file=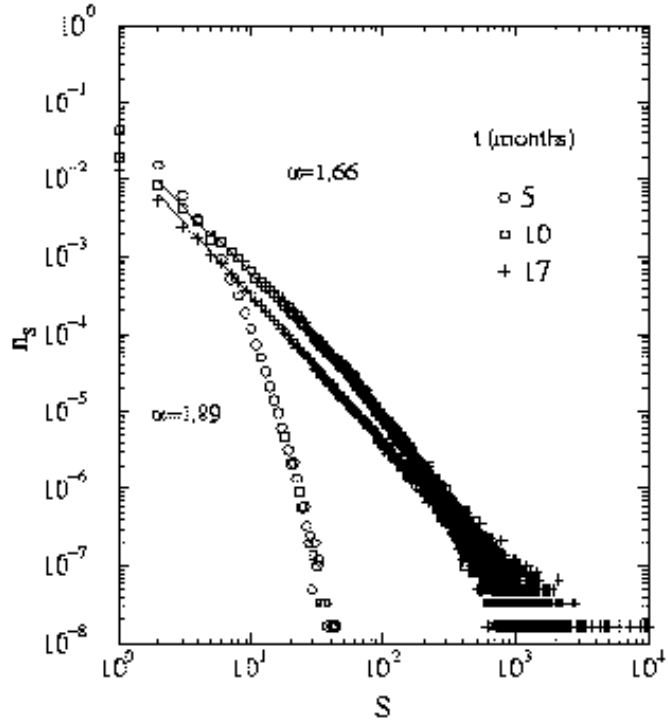,width=9cm,angle=0}}
\caption{Simulated cluster size distribution function $n_s(t)$ for the grove SJ71 at various observation times. 
The straight line represent the best fit to the data; its slope gives the exponent $\alpha$ describing }
\label{fig9}
\end{figure}

\begin{figure}[f]
\centerline{\epsfig{file=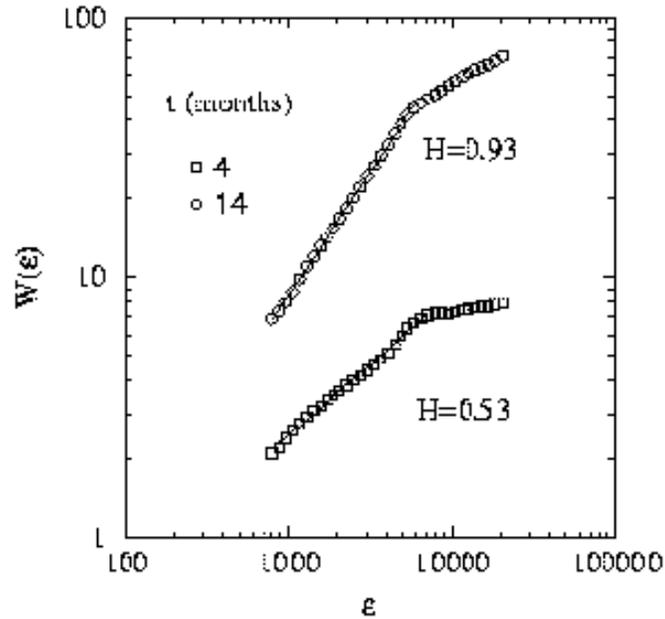,width=9cm,angle=0}}
\caption{Typical log-log plot of $W(\epsilon) \times \epsilon$ corresponding to the simulated infection profiles 
for the grove SJ71 at two observation times. The straight line represent the best fit to the data; its slope gives the 
roughness exponent $H$ describing the profile. The data correspond to an average over $20$ different realizations of CA evolution.}
\label{fig10}
\end{figure}

In figure \ref{fig8}, it is shown a simulated temporal sequence of CVC incidence maps qualitatively similar 
to the observed spatial patterns of CVC (see figure \ref{fig2}). From such incidence maps, one can determine 
the dynamic infected cluster size distribution function, $n_s(t)$, and the roughness exponent $H$ 
characterizing the self-affinity of CVC infection profiles. Figures \ref{fig9} and \ref{fig10} show, respectively, the 
cluster size distribution functions $n_s(t)$, and typical log-log plots of $W(\epsilon) \times \epsilon$ 
corresponding to the simulated infection maps for the grove SJ71 at various observation times. Both, 
the power-law decay of the infected cluster size distribution functions and the roughness exponents 
characterizing the spatial disease patterns, indicate the presence of long-range correlations in the CVC 
development. Roughness exponents greater than $1/2$ mean that in the neighborhood of a diseased plant 
the probability to find another infected tree increases. A significative aggregation of diseased plants 
suggest that the pathogen predominantly spreads from plant to plant. Since our CA model permits fast simulations 
of many large samples of artificial CVC pathosystems under various epidemiological contexts, the numerical 
values for the exponents $\alpha$ and $H$ can be determined with a reliable statistical precision difficultly 
attained in actual field observations.

It is important to emphasize that a simple random walk distribution for the inoculative vector flights 
constrained to a local neighborhood of radius $r(t)$ around each diseased plant, also  generates progress 
curves in good agreement with the field data. However, the resulting CVC incidence maps are clearly 
distinct from the observed ones. Random flights of inoculative vectors produce spots of diseased plants 
artificially isolated in space, which grow up to merge with other infected clusters. In contrast, 
a L\'evy distribution permits rare long range vector flights which appears to be an essential feature 
to explain the scale invariance observed in CVC spreading. Indeed, conventional random walks used 
to model foraging behavior in biology~\cite{Berg} predict a Poisson instead of a scale-invariant 
power-law distribution. Thus, our results suggest that the inoculative sharpshooter leafhoppers 
($\sim 1 \, cm$ in size) perform long flights of random foraging, searching for non-stressed plants 
with new vegetative growth unpredictably dispersed over several square kilometers. A possible 
explanation is that for insects operating in swarms or flocks comprised of $N$ walkers, L\'evy 
flight search patterns (for which $Nt$ sites are visited after $t$ steps) are much more efficient 
than Brownian walk foraging patterns (for which only $tln(N/lnt)$ distinct sites are visited)~\cite{Larralde}. 
It is  interesting to mention that L\'evy flight search strategies are also observed in albatrosses~\cite{Viswanathan}. 

\section{Conclusions}

In this study the spatio-temporal analysis of CVC spreading was carried out. The shape of the observed 
CVC progress curves was double sigmoid best fitted by five parameters generalized logistic function. 
This means that CVC is a polyciclic disease in which a phase of rapid progress alternates with another 
one of almost paralyzes. In addition, both the power-law decay of the infected cluster size distribution 
functions and the roughness exponents characterizing the spatial disease patterns, indicate the presence 
of long-range correlations in the CVC development.

In order to understand the basic mechanisms by which the features discussed above emerge, a 
CA model was proposed. It takes into account the motility of sharpshooter vectors, the hydric 
and nutritional level of plant stress, as well as seasonal climatic effects. Varying the CA 
parameters controlling these factors, a good agreement among simulational and all the observational 
data was achieved, suggesting that the actual relevant mechanisms of CVC spreading were really captured 
and evidenced by the evolution rules of the proposed CA model. Therefore, our model suggest that 
the average number $n_v(t)$ of L\'evy flights performed by the sharpshooter vectors as well as plant 
stress, described by the probability $p_{st}$, are the most fundamental parameters determining the 
aspects of CVC spreading. Also, all of them are affected by seasonal variations.

\section{Acknowledgements}

This work was partially supported by the CNPq, and FAPEMIG - Brazilian agencies.


\vspace{1cm}

{\bf Table 1} Coefficients of determination $R^2$ and estimated values of 
the five parameters describing the generalized logistic functions used to 
fit the observed data of CVC progress in four of the studied groves. 

\begin{table}[ht]
\vspace{7pt} \centering
\begin{tabular}{|c|c|c|c|c|c|c|}
\hline\hline
{\rm Parameters} & $p_1$ & $p_2$ & $p_3$ & $p_4$ & $p_5$ & $R^2$ \\ \hline\hline
{\rm SJ01} & 0.223 & -6.000 & 1.647 & -0.158 & 0.005 & 0.989 \\ \hline
{\rm SJ67} & 0.908 & -6.414 & 1.665 & -0.153 & 0.005 & 0.996 \\ \hline
{\rm SJ71} & 1.000 & -4.854 & 1.307 & -0.126 & 0.004 & 0.989 \\ \hline
{\rm SJ75} & 0.526 & -6.638 & 1.396 & -0.120 & 0.004 & 0.993 \\ \hline\hline
\end{tabular}
\end{table}


{\bf Table 2} CA parameters used to simulate CVC progress curves corresponding to the observed data for four groves. 

\begin{table}[ht]
\vspace{7pt} \centering
\begin{tabular}{|c|c|c|c|c|c|}
\hline\hline
{\rm Parameters} & $\mu$ & $a$ & $n_0$ & $p_{inf}$ & $p_{min}$ \\ \hline\hline
{\rm SJ01} & 4 & 0.68 & 5 & 0.0002 & 0.001 \\ \hline
{\rm SJ67} & 4 & 0.68 & 5 & 0.04 & 0 \\ \hline
{\rm SJ71} & 4 & 0.68 & 5 & 0.06 & 0.35 \\ \hline
{\rm SJ75} & 4 & 0.68 & 5 & 0.009 & 0 \\ \hline\hline
\end{tabular}
\end{table}


\end{multicols}
\widetext

\newpage

\end{document}